\documentclass[%
reprint,
superscriptaddress,
 amsmath,amssymb,
 aps,
prl,
]{revtex4-1}

\usepackage{graphicx}
\usepackage{dcolumn}
\usepackage{bm}
\usepackage{amsmath,amssymb}
\usepackage{xcolor}
\usepackage{color,soul}

\begin{document}

\title{Disentangling spin-orbit coupling and local magnetism  \\
 in a quasi-2D electron system}

\author{Xinxin Cai}
\affiliation{School of Physics and Astronomy, University of Minnesota, MN 55455, USA}
\author{Yilikal Ayino}
\affiliation{School of Physics and Astronomy, University of Minnesota, MN 55455, USA}
\author{Jin Yue}
\affiliation{Department of Chemical Engineering and Materials Science, University of Minnesota, MN 55455, USA}
\author{Peng Xu}
\affiliation{Department of Chemical Engineering and Materials Science, University of Minnesota, MN 55455, USA}
\author{Bharat Jalan}
\affiliation{Department of Chemical Engineering and Materials Science, University of Minnesota, MN 55455, USA}
\author{Vlad S. Pribiag}
\email{vpribiag@umn.edu}
\affiliation{School of Physics and Astronomy, University of Minnesota, MN 55455, USA}

\date{\today}

\begin{abstract}
Quantum interference between time-reversed electron paths in two dimensions leads to the well-known weak localization correction to resistance. If spin-orbit coupling is present, the resistance correction is
negative, termed weak anti-localization (WAL). Here we report the observation of WAL coexisting with exchange coupling between itinerant electrons and localized magnetic moments. We use low-temperature magneto-transport measurements to investigate the quasi-two-dimensional, high-electron-density interface
formed between SrTiO$_3$ (STO) and the anti-ferromagnetic Mott insulator NdTiO$_3$ (NTO). As the magnetic field angle is gradually tilted away from the sample normal, the data reveals the interplay between strong $k$-cubic Rashba-type spin-orbit coupling and a substantial magnetic exchange interaction from local magnetic regions. The resulting quantum corrections to the conduction are in excellent agreement with existing models and allow sensitive determination of the small magnetic moments (22 $\mu_B$ on average), their magnetic anisotropy and mutual coupling strength. This effect is expected to arise in other 2D magnetic materials systems. 
\end{abstract}

\maketitle

Quantum interference of time-reversed electron paths in a diffusive conductor gives rise to weak localization corrections to the conductance. In the presence of spin-orbit coupling (SOC), the interference becomes destructive, resulting in enhanced conductance near zero magnetic field and hence positive magneto-resistance (MR), known as weak anti-localization (WAL), which can be analyzed to extract SOC parameters \cite{hikami_spin-orbit_1980,altshuler_anomalous_1981,iordanskii_weak_1994}. 
Conventional WAL occurs in two-dimensional samples with no intrinsic magnetism, subject to a weak perpendicular magnetic field. In contrast, here we investigate experimentally a distinct effect: the interplay between SOC and strong magnetic exchange, and show that WAL can provide a sensitive quantitative probe not only of SOC, but also of local magnetic properties.

Our experimental system consists of metallic interfaces 
between STO and NTO \cite{xu_stoichiometry-driven_2014,xu_quasi_2016, xu_predictive_2016}. Interfaces between two complex oxides \cite{ohtomo_high-mobility_2004,takizawa_photoemission_2006,moetakef_electrostatic_2011,xu_stoichiometry-driven_2014} can host a quasi-two-dimensional conducting electron gas which exhibits a rich variety of phenomena \cite{sulpizio_nanoscale_2014}, ranging from superconductivity \cite{reyren_superconducting_2007,caviglia_electric_2008,biscaras_two-dimensional_2010} 
to strong spin-orbit coupling \cite{ben_shalom_tuning_2010,caviglia_tunable_2010,bal_strong_2018} 
and magnetism \cite{brinkman_magnetic_2007,li_coexistence_2011,bert_direct_2011,dikin_coexistence_2011,moetakef_carrier-controlled_2012,joshua_gate-tunable_2013,anahory_emergent_2016,bal_strong_2018,ayino_ferromagnetism_2018}. 
NTO is an anti-ferromagnetic (AF) Mott-Hubbard insulator, featuring long-range magnetic
ordering on the Ti$^{3+}$ sublattice with a N$\acute{\text{e}}$el temperature of $\sim90$~K \cite{amow_structural_1996,sefat_effect_2006}. The NTO/STO interfaces in this study are grown using the hybrid molecular beam epitaxy technique (hMBE) \cite{jalan_molecular_2009} that ensures excellent control over stoichiometry for the growth of complex oxide thin films \cite{xu_stoichiometry-driven_2014}. The quasi-two-dimensional electron gas (q2DEG) resides on the STO side of the interface and has ultra-high carrier densities that can, for reference, be one or two orders of magnitude higher than typically seen in LaAlO$_3$ (LAO)/STO \cite{xu_quasi_2016}. Here, we focus on a hetero-interface with layer thicknesses STO(8~u.c.)/NTO(2~u.c.)/STO(8~u.c.)/(La,Sr)(Al,Ta)O$_3$ (LSAT)(001) (substrate). The extra STO capping layer is grown to protect NTO from degradation due to oxygen absorption in the air \cite{xu_predictive_2016}. We note that the electrons which accumulate at the STO-on-NTO type interfaces tend to have very low electron mobility and exhibit insulating behavior at low temperatures \cite{xu_quasi_2016, ayino_ferromagnetism_2018}. Our measurement and data analysis in this work treat the sample as a single quasi-2D electron system consistently and show no effects arising from parallel conduction at the top interface.

\begin{figure}[t]
\includegraphics{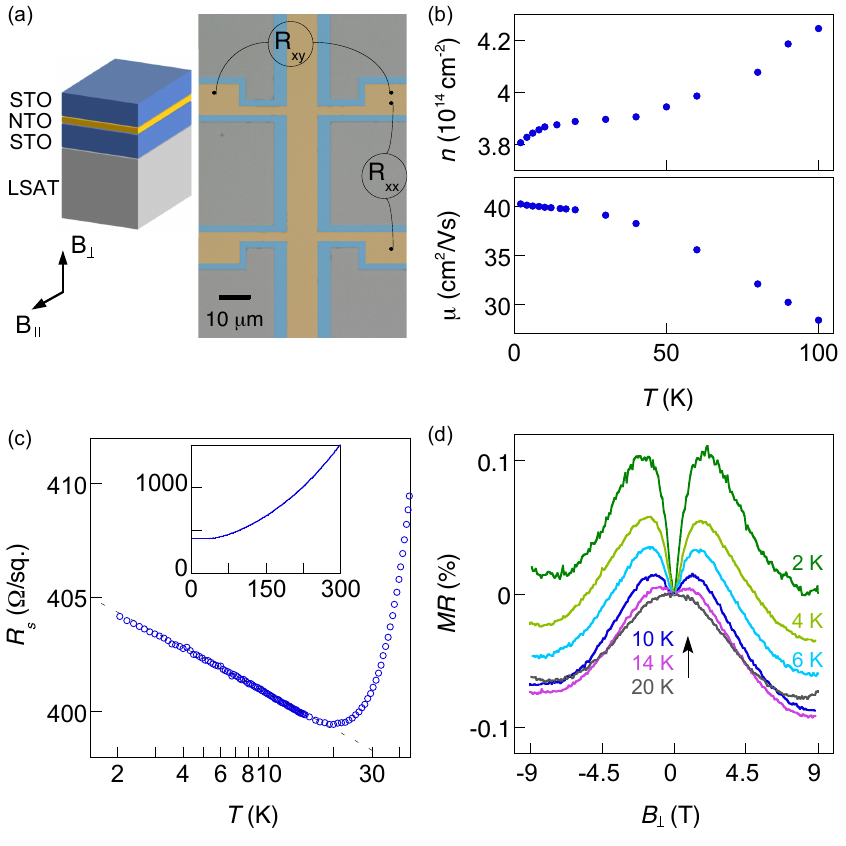}
 \caption{\label{fig1} (Color online). (a) Left: A schematic of the capped STO(8~u.c.)/NTO(2~u.c.)/STO(8~u.c.)/LSAT (001) heterostructure. 
The directions of applied fields are indicated with respect to the heterointerface. Right: False-color optical image of a typical Hall-bar sample prepared on the heterostructure. The etched regions are indicated in blue.
(b) The temperature dependence of carrier density $n$ and Hall mobility $\mu$.
(c) The sheet resistance $R_s$ as a function of temperature in a logarithmic scale. The insets are the corresponding $R_s$ measured up to the room temperature.
(d) Magnetoresistance at various temperatures, MR\%=$\frac{R_{s}-R_{s,B=0}}{R_{s,B=0}}\times100\%$.}
\end{figure}

To facilitate magneto-transport measurements and analysis, $10\times20$ and $10\times40~\mu$m$^2$ Hall-bar devices are etched by Ar ion milling (right panel in Fig.~1a). Temperature and magnetic field are controlled in a 9~T Quantum Design PPMS system at $T$ down to $2$~K. A rotational sample holder is used to apply fields at various angles with respect to the sample plane. 
Four-terminal resistance is measured using DC currents $\leq$0.5~$\mu$A. 
The temperature dependence of electron density $n$ and mobility $\mu$ 
of the heterointerface,
obtained from the longitudinal resistance and Hall effect data, are presented in Fig.~1b.
The heterointerface is metallic and shows a logarithmic-like increase in resistance with decreasing temperature below 20~K (Fig.~1c).
The sample magnetoresistance (MR) as a function of perpendicular field $B_\bot$ is measured at various temperatures within the log-$T$ regime; as shown in Fig.~1d, sharp positive MR is clearly seen around zero field for $2$~K, which we attribute to quantum interference in the presence of SOC (WAL).

The magneto-conductance correction $\Delta\sigma(B_\bot)$ due to WAL has the following form \cite{iordanskii_weak_1994,minkov_weak_2004}: 
\begin{equation}
\sigma(B_\bot)-\sigma(0)=\sigma_0\left[F_t\left(\frac{B_\phi}{B_\bot},\frac{B_{so}}{B_\bot}\right)-F_s\left(\frac{B_\phi}{B_\bot}\right)\right],
\end{equation}
where $\sigma_0=e^2/\pi h$, and $B_\phi$ and $B_{so}$ are the effective fields characterizing the phase and spin relaxation of the electrons, respectively.
The function $F_t$ describes the positive contribution from the interfering electron waves in the triplet state with a total spin of $J=1$,
while the singlet state ($J=0$) contributes a negative correction, described by $-F_s$. 
For 2D structures with inversion symmetry breaking, the specific expressions for the functions were derived
  by Iordanskii, Lyanda-Geller, and Pikus (ILP), and incorporate the mechanisms of spin relaxation arising from both the $k$-linear and  $k$-cubic spin-orbit splitting of electron spectra \cite{iordanskii_weak_1994}. 
Importantly, the appearance of a local maximum in the WAL MR in 2D structures, as shown in Fig.~1d, is an indication that the dominant mechanism of spin relaxation is the Dyakonov-Perel type, arising from spin-splitting, rather than the Elliott-Yafet mechanism due to spin-flip scattering by impurities \cite{knap_weak_1996}.

\begin{figure}[t]
\includegraphics{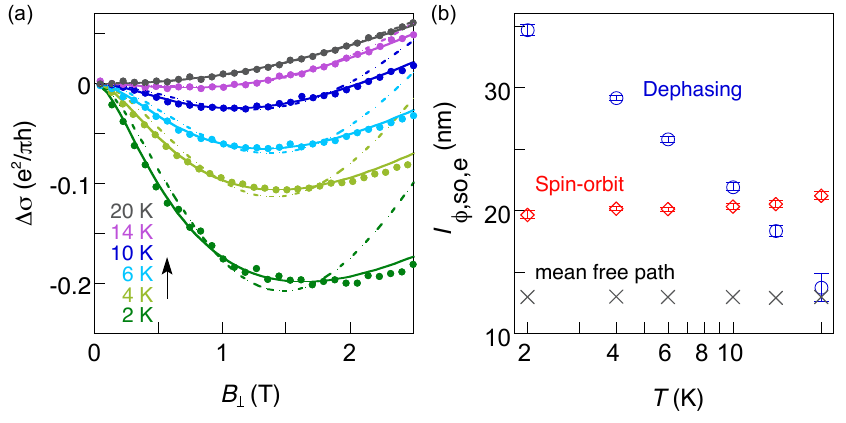}%
 \caption{\label{fig:WAL} (Color online). 
 (a) Conductance correction $\Delta \sigma$ (dots) in units of $e^2/\pi h$, derived from the MR measurements by subtracting the $B_\bot^2$ background. Theoretical fits to the ILP theory including only the $k$-linear spin splitting (dashed line) or $k$-cubic spin splitting (solid line). The data are accurately described only by the $k$-cubic model. 
(b) The extracted phase coherence length $l_\phi$ and the spin precession length $l_{so}$ for the $k$-cubic case, plotted alongside the mean free path $l_e$, as a function of temperature. }
\end{figure}

In order to obtain quantitative information about the SOC we next analyze the magneto-transport data for perpendicular applied fields. Fig.~2a shows the measured $\Delta\sigma(B_\bot)$ in units of $\sigma_0$, obtained by subtracting a classical positive $B_\bot^2$ background ($\Delta\sigma\approx -\frac{1}{R_s}\left[\frac{\Delta R_{s}}{R_{s}}-(\mu B_\bot)^2\right]$).
We fit the $\Delta\sigma$ curves to the ILP model using $B_\phi$ and $B_{so}$ as the variables \cite{iordanskii_weak_1994,Supinfo}. The fits are restricted to the low-field regime in the diffusive approximation for $B_\bot<B_e=\hbar/2el_{e}^2$ ($\approx2$~T in the present system). In many systems, the $k$-linear Rashba term is dominant. Interestingly however, as can be seen in Fig.~2a, our data are well reproduced by considering the $k$-cubic splitting only, and deviate significantly from the model with only $k$-linear splitting. 
We note that the crystal structures of epitaxially grown STO and NTO are both centrosymmetric \cite{jalan_molecular_2009,jeong_new_2016}, leading to no Dresselhaus terms. The fitting results indicate that the dominant form of SOC is likely cubic Rashba, allowed by the interfacial asymmetry of the q2DEG and possibly associated with the Ti 3$d$ orbitals in STO
\cite{nakamura_experimental_2012}. 

Additional information about the transport properties of the interface can be obtained from these data. The phase coherence length $l_\phi$ and the spin-orbit length $l_{so}$
obtained from the fits to the $k$-cubic model
are plotted as a function of temperature in Fig.~2b, derived using the relation $B_{i}=\hbar/4el_{i}^2$, $_{i=so,\phi}$.
$l_{so}$ is relatively independent of $T$, as expected, and remains around 20~nm, corresponding to a large spin-orbit field, $B_{so}\sim0.4$~T. To estimate the spin relaxation rate $\tau_{so}^{-1}$ and the spin splitting $\Delta$, we apply the relations $l_{so}=\sqrt{D \tau_{so}}$, 
$B_{so}=\Delta^2\tau_e/2eD\hbar$,
$D=v_F^2\tau_e/2$, $v_F=\hbar k_F/m^*$ and $\tau_e=\mu m^*/e$, where $D$ is the diffusion constant, $v_F$ the Fermi velocity, $\tau_e$ is the elastic scattering time,  and $k_F=\sqrt{2 \pi n}$ is the Fermi wave vector. Taking the effective mass to be $m^*=0.8m_e$ \cite{mattheiss_effect_1972-1}, we have $\tau_{so}^{-1}\sim$11~(ps)$^{-1}$ and a large Rashba spin splitting, $\Delta\sim$0.012~eV. 
Fig.~2b also shows the mean free path $l_e$ obtained from Hall measurements. The localization theory is applicable at temperatures no higher than $\sim20$~K for $l_\phi>l_e$. This boundary is consistent with the onset temperature at which the log-$T$ increase in resistance emerges (Fig.~1c). 

We next move beyond the standard WAL analysis to investigate the influence of a magnetic field parrallel to the sample plane, $B_{||}$, on the quantum interference of electrons. The magneto-conductance correction $\Delta\sigma$ as a function of $B_{||}$ is presented in Fig.~3a. Intriguingly, we observe pronounced negative $\Delta\sigma(B_{||})$ 
for the entire temperature range, with a sharp drop at low $B_{||}$ and a gradual decrease as $B_{||}$ is further increased. Decreasing $T$ enhances the overall magnitude of $\Delta\sigma$.

It is known that a parallel field can lead to negative $\Delta\sigma$ due to the Zeeman interaction in the presence of SOC \cite{maekawa_magnetoresistance_1981,malshukov_magnetoresistance_1997,zumbuhl_spin-orbit_2002}. 
The effect of the Zeeman interaction is to further suppress the singlet state of the interfering electrons, resulting in additional dephasing. This additional singlet dephasing contribution, $\Delta_\phi$, is described by \cite{minkov_weak_2004,malshukov_magnetoresistance_1997}:
\begin{equation}
\Delta_\phi(B_{||})=\frac{(g\mu_BB_{||})^2}{(4eD)^2B_{so}}.
\end{equation}
$\Delta\sigma(B_{||})$ has the following form based on the ILP theory  \cite{Supinfo,minkov_weak_2004}: 
\begin{equation}
\sigma(B_{||})-\sigma(0)=-\frac{\sigma_0}{2}\ln\left(1+\frac{\Delta_\phi}{B_\phi}\right).
\end{equation}
$B_{so}$ and $B_\phi$ are the low-field values extracted from the $\Delta\sigma(B_\bot)$ fits for $k$-cubic spin splitting (Fig.~2).
An estimation of the Zeeman effect 
is shown by the dashed curves in Fig.~3a, assuming 
reasonable values of $m^*$ and $g$ based on the analysis below. As can be seen from the plot, the Zeeman effect due to the applied field alone is far too weak to account for the data.

In a quasi-2D system, the parallel field may also influence the localization correction via the non-vanishing orbital motion in the $z$ direction.
Previous work on this effect includes studies of
the role of micro-roughness in 2D structures \cite{mathur_random_2001}, subband intermixing \cite{meyer_quantum_2002}, and tunneling between parallel quantum wells \cite{raichev_weak-localization_2000}. 
Importantly, unlike the Zeeman interaction, these mechanisms would affect the phase coherence of the singlet and triplet states indistinguishably and would lead to an additional dephasing term ($\Delta_\phi'$) that would depend primarily on $B_{||}^2$ for each of the two spin states.
As a result, such mechanisms would result in a further positive correction to $\Delta\sigma$ due to the triplet contribution, which would be of opposite sign to the observed $\Delta\sigma$ 
and to the Zeeman correction. Therefore, based on the ILP model, the overall $\Delta\sigma$ in the presence of both orbital and Zeeman effects would be weakly negative at low fields and turning positive at high fields \cite{Supinfo,minkov_weak_2004}.
Such nonmonotonic $B_{||}$-dependence of $\Delta\sigma$ is not seen in the data at any temperature, 
indicating that the orbital effects of $B_{||}$ are insignificant.

\begin{figure}[t]
\includegraphics{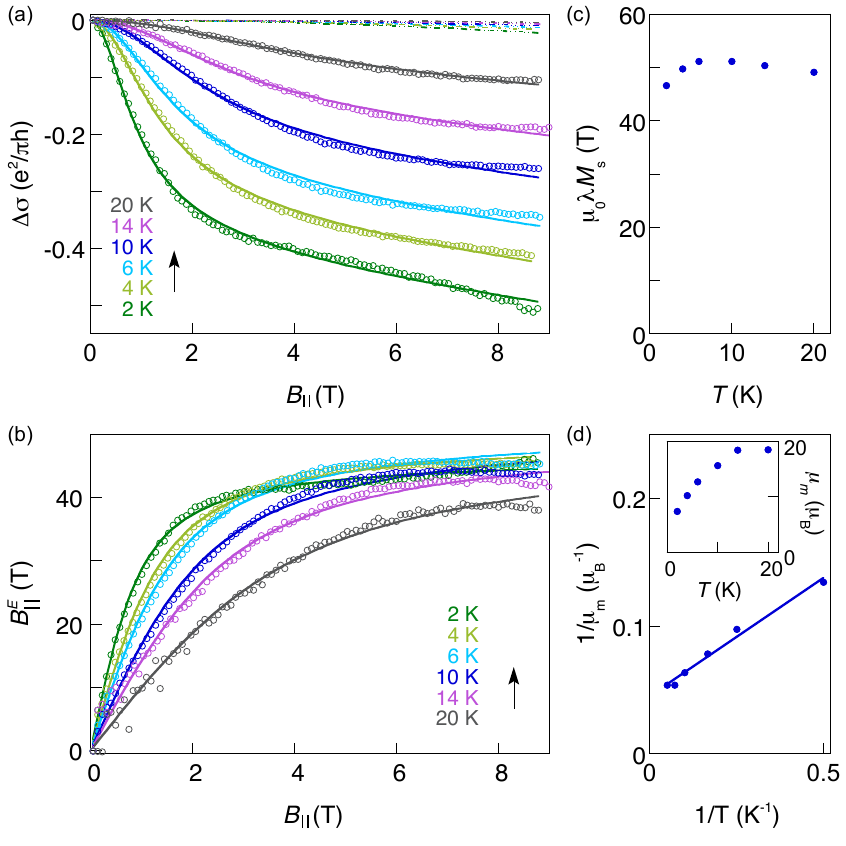}
 \caption{\label{Fig. 1} (Color online). 
 (a) Conductance correction $\Delta \sigma$ in units of $e^2/\pi h$ as a function of the parallel field $B_{||}$: 
 experimental data (dots) and theoretical fits incorporating the exchange field (solid lines). The contribution of the Zeeman effect due to the bare applied field only, excluding  the exchange field, is shown by dashed lines at the top. 
$gm^*/m_e=1.5$ is assumed for the theoretical fits.
 (b) Exchange field $B_{||}^E$ as a function of $B_{||}$ derived from the $\Delta \sigma(B_{||})$ data using Eq.~(4) (dots) and from theoretical fits to the Langevin function (solid lines). 
 (c) The saturation exchange field $\mu_0\lambda M_s$ extracted from the fits. (d) The inverse of the apparent local moment 1/$\mu_m$ as function of 1/$T$. Inset: extracted values of $\mu_m$ $vs.$~$T$.}
\end{figure}

Having excluded orbital effects and the role of the bare applied field, we propose that the unusual observed $\Delta\sigma(B_{||})$ behavior is associated with the magnetic structure of the sample, specifically the effect of the magnetic exchange interaction on conduction electron spins at the interface. 
The exchange interaction can be represented by an effective exchange field $B^{E}$, which couples only to electron spins but, importantly, has no direct effect on the orbital motion \cite{ohandley_modern_2000}.
It affects $\Delta\sigma$ through the Zeeman term in similar fashion as a large magnetic field,
leading to quick dephasing of the singlet state \cite{dugaev_weak_2001}. Therefore, to better examine the role of the exchange interaction, we utilize Eqs.~(2) and (3), replacing the applied $B_{||}$ by a total effective field in the plane, $B_{||}^t=B_{||}+B_{||}^{E}$. Eq.~(2) becomes:
\begin{equation}
\Delta_\phi(B_{||})=\frac{(g\mu_B)^2(B_{||}+B_{||}^{E})^2}{(4eD)^2B_{so}}.
\end{equation}
The value of $\Delta_\phi$ depends on the product of $g$ and $m^*$. We examine the slope of $\sqrt{\Delta_\phi}$~vs.~$B_{||}$ in the high-$B_{||}$ region of the 2~K data, where we assume $B^{E}_{||}$ has reached saturation, and obtain an estimation of $gm^*/m_e\sim1.5$. Accordingly, $B_{||}^E$ as a function of $B_{||}$ is derived from the $\Delta\sigma(B_{||})$ data for all temperatures and plotted in Fig.~3b. Importantly, the $B_{||}$-dependence of $B^{E}_{||}$ is well described by the Langevin function L(x), characteristic for an ensemble of superparamagnets. Each nanoscale superparamagnetic region consists of a group of spins with local ferromagnetic order, which collectively behave as a large classical paramagnetic moment \cite{ohandley_modern_2000}, as previously established in NTO/STO interfaces \cite{ayino_ferromagnetism_2018}. The magnetism within each nano-region is possibly associated with canted spins of the antiferromagnetic NTO adjacent to the interface, due to the Dzyaloshinskii-Moriya interaction \cite{ayino_ferromagnetism_2018,sefat_effect_2006}.

To quantitatively analyze the magnetic properties of the interface, we apply the standard relation $B^{E}=\mu_0\lambda M=\mu_0\lambda M_s\text{L}(\mu_m B/k_B T)$, where $\lambda$ is the coefficient characterizing the effective exchange interaction between electrons and local moments, $M_s$ is the saturation magnetization, and $\mu_m$ is the moment of a single magnetic region. 
We note that the localization theory is not valid at very high $B_{||}$ when the combined exchange and Zeeman interaction is large enough to mix the singlet and the triplet states. As a result, the above analysis is limited to the condition $g\mu_BB^t_{||}<\hbar/\tau_{so}$ \cite{malshukov_magnetoresistance_1997}, that is $\Delta_\phi<B_{so}$. This condition is found to hold for the data in the entire measurement range shown in Fig.~3a. Using $\mu_0\lambda M_s$ and $\mu_m$ as two variables, the $\Delta\sigma(B_{||})$ and $B_{||}^{E}$ data are very well reproduced by the fits incorporating the Langevin function into Eqs.~(3) and (4), as demonstrated in Figs.~3a and 3b respectively. 

The extracted value of $\mu_m$ shows an artificial decrease with decreasing temperature (inset in Fig.~3d) despite little change in the value of the saturation magnetization (Fig.~3c). However, a closer look reveals that the inverse of the apparent moment, $1/\mu_m$, changes linearly with the inverse of the temperature (Fig.~3d). This behavior is consistent with the scenario of weakly interacting superparamagnets, where the true magnetic moment $\mu^*_m$ follows the relation $1/\mu_m=1/\mu^*_m(1+T^*/T)$ and $T^*$ characterizes the energy scale of the dipole-dipole interaction \cite{allia_granular_2001}.
From the linear fit in Fig.~3d, we obtain the average magnetic moment of a single nanoscale magnetic region to be $\mu^*_m=22 \mu_B$, and $T^*=4.1$~K, corresponding to an rms dipolar energy $k_BT^*\sim0.35$~meV. 

The exchange field $B^E$ in the above analysis is an averaged effect for short-range coupling of conduction electrons to the nanoscale magnetic regions. Above the blocking temperature, the local exchange field of each nanoscale region jumps among different stable orientations as a result of thermally-induced superparamagnetic fluctuations, and these fluctuations gradually become polarized in the direction of the parallel applied field as its magnitude is increased. Many such magnetic regions, each of which has a randomly oriented exchange field, are present in the system and overlap with the closed time-reversed paths of the electrons that lead to WL/WAL. The net effect of the superparamagnetic fluctuations on all the time-reversed electron paths is an effective exchange field that is well fitted by the Langevin function (Fig. 3(b)). Since an exchange field acts only on electron spins, it has no effect on the orbital motion and thus does not lead to conventional WL/WAL.

\begin{figure}[t]
\includegraphics{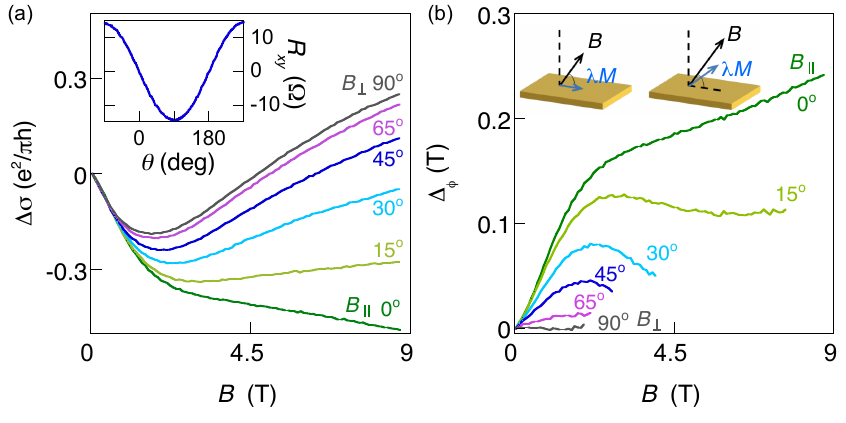}
 \caption{\label{Fig. 1} (Color online). 
 (a) Conductance correction $\Delta \sigma$ in units of $e^2/\pi h$ as a function of the applied field $B$ at an angle $\theta$ from the interface plane, obtained by subtracting the $(\mu B$sin($\theta))^2$ background from the MR data. Inset: 
Hall resistance $R_{xy}$ as a function of $\theta$ at $B=9$~T.
 (b) Calculated $\Delta_\phi$ as a function of $B$. Inset: schematics showing the exchange field lying in the easy-plane at small $B$ (left) and developing an out-of-plane component at larger $B$ (right).}
\end{figure}

Next, we investigate the angular-dependence of the quantum interference and the anisotropy of the magnetic exchange. To this end, we measure
the MR 
when the applied field $B$ forms a tilt angle $\theta$ with respect to the sample plane (Fig.~4a). The Hall effect shows a perfect sin$(\theta)$-dependence, as expected (inset in Fig.~4a).
The parallel component of $B$ adds an additional dephasing term $\Delta_\phi$ for the singlet state and therefore Eq.~(1) becomes:
\begin{align}
&\sigma(B_\bot,B_{||})-\sigma(0,B_{||}) \nonumber \\
&=\sigma_0\left[F_t\left(\frac{B_\phi}{B_\bot},\frac{B_{so}}{B_\bot}\right)-F_s\left(\frac{B_\phi+\Delta_\phi}{B_\bot}\right)\right],
\end{align}
Thus, the total magneto-conductance correction $\Delta\sigma(B_\bot,B_{||})=\sigma(B_\bot,B_{||})-\sigma(0,0)$ is described by the sum of Eq.~(3) and Eq.~(5). Using these equations, we extract the dependence of $\Delta_\phi$ as a function of $B$ for $B_\bot<B_e$, in the diffusive regime  (Fig.~4b). Interestingly, the evolution of $\Delta_\phi$ with $B$ supports the scenario that the magnetization has an easy-plane anisotropy, as expected for a thin magnetic film.  For a fixed angle, the value of $\Delta_\phi$ is determined by the in-plane component of the total field (Eq.~4). The magnetization primarily lies in the plane, leading to an increasing $\Delta_\phi$ at low fields. At large enough values, the applied field overcomes the in-plane anisotropy and pulls the magnetization out of the plane, and therefore the in-plane component of the exchange field drops, leading to a decrease of $\Delta_\phi$ (shown schematically in Fig.~4b). Magneto-transport behaviors similar to those presented above have been observed in other Hall-bar devices \cite{Supinfo}.  Moreover, we see no dependence on the direction between applied field and current.

In conclusion, by tilting the magnetic field into the sample plane, we observed a distinct coherent electron interference effect due to the interplay between the SOC and the magnetic exchange interaction arising from local magnetism. 
The presence of interfacial ferromagnetic order within each local region gives rise to a substantial exchange field (tens of Tesla) that couples to the conduction electron spins. The effective exchange field leads to a dramatically enhanced Zeeman effect, which contributes to the localization in the presence of SOC.  
This is qualitatively different from the commonly-studied case of magnetic impurity scattering, which leads to the dephasing of electron waves and weakens the WAL/WL \cite{hikami_spin-orbit_1980,maekawa_magnetoresistance_1981}. In that case, the magnetic impurities are small local paramagnets with no spin correlations or magnetic order, and therefore allow the conduction electrons to exchange spin angular momentum during the scattering process, leading to spin-flip dephasing.

The effect we report here is sensitive to the exchange field and thus also serves as a probe of the local magnetic moment, since $B^{E}\propto~M$. In all, our approach, using closed time-reversed electron paths, allows a highly sensitive method for determining magnetic moments as small as $\sim22~\mu_B$ on average and provides detailed information about their collective anisotropy and mutual couplings. It is a broadly-accessible approach based on magnetotransport measurements and does not require any magnetometery equipment. It is also interesting to note that the most common magnetometry techniques, such as SQUID or Kerr effect typically lack the necessary sensitivity to detect and quantify such small moments, particularly in the superparamagnetic regime.   
While the effect is observed here at the epitaxial interface between NTO and STO, it is expected to be relevant for other ferro-magnetic and anti-ferromagnetic 2D systems.

Acknowledgements: We thank Z.~Jiang and B.~Yang for assistance with PPMS measurements, and P.~Crowell, L.~Zhao and Y.~Iguchi for valuable discussions. This work was supported primarily by the Office of Naval Research under Award No. N00014-17-1-2884. Film growth and structural characterizations were funded by the U.S. Department of Energy through the University of Minnesota Center for Quantum Materials, under Grant No. DE-SC-0016371. Portions of this work were conducted  in the Minnesota Nano Center, which is supported by the National Science Foundation through the National Nano Coordinated Infrastructure Network (NNCI) under Award Number ECCS-1542202. Sample structural characterization was carried out at the University of Minnesota Characterization Facility, which receives partial support from NSF through the MRSEC program under Award No. DMR-1420013.


\bibliography{Refs}

\begin{thebibliography}{44}%
\makeatletter
\providecommand \@ifxundefined [1]{%
 \@ifx{#1\undefined}
}%
\providecommand \@ifnum [1]{%
 \ifnum #1\expandafter \@firstoftwo
 \else \expandafter \@secondoftwo
 \fi
}%
\providecommand \@ifx [1]{%
 \ifx #1\expandafter \@firstoftwo
 \else \expandafter \@secondoftwo
 \fi
}%
\providecommand \natexlab [1]{#1}%
\providecommand \enquote  [1]{``#1''}%
\providecommand \bibnamefont  [1]{#1}%
\providecommand \bibfnamefont [1]{#1}%
\providecommand \citenamefont [1]{#1}%
\providecommand \href@noop [0]{\@secondoftwo}%
\providecommand \href [0]{\begingroup \@sanitize@url \@href}%
\providecommand \@href[1]{\@@startlink{#1}\@@href}%
\providecommand \@@href[1]{\endgroup#1\@@endlink}%
\providecommand \@sanitize@url [0]{\catcode `\\12\catcode `\$12\catcode
  `\&12\catcode `\#12\catcode `\^12\catcode `\_12\catcode `\%12\relax}%
\providecommand \@@startlink[1]{}%
\providecommand \@@endlink[0]{}%
\providecommand \url  [0]{\begingroup\@sanitize@url \@url }%
\providecommand \@url [1]{\endgroup\@href {#1}{\urlprefix }}%
\providecommand \urlprefix  [0]{URL }%
\providecommand \Eprint [0]{\href }%
\providecommand \doibase [0]{http://dx.doi.org/}%
\providecommand \selectlanguage [0]{\@gobble}%
\providecommand \bibinfo  [0]{\@secondoftwo}%
\providecommand \bibfield  [0]{\@secondoftwo}%
\providecommand \translation [1]{[#1]}%
\providecommand \BibitemOpen [0]{}%
\providecommand \bibitemStop [0]{}%
\providecommand \bibitemNoStop [0]{.\EOS\space}%
\providecommand \EOS [0]{\spacefactor3000\relax}%
\providecommand \BibitemShut  [1]{\csname bibitem#1\endcsname}%
\let\auto@bib@innerbib\@empty
\bibitem [{\citenamefont {Hikami}\ \emph {et~al.}(1980)\citenamefont {Hikami},
  \citenamefont {Larkin},\ and\ \citenamefont
  {Nagaoka}}]{hikami_spin-orbit_1980}%
  \BibitemOpen
  \bibfield  {author} {\bibinfo {author} {\bibfnamefont {S.}~\bibnamefont
  {Hikami}}, \bibinfo {author} {\bibfnamefont {A.~I.}\ \bibnamefont {Larkin}},
  \ and\ \bibinfo {author} {\bibfnamefont {Y.}~\bibnamefont {Nagaoka}},\ }\href
  {\doibase 10.1143/PTP.63.707} {\bibfield  {journal} {\bibinfo  {journal}
  {Progress of Theoretical Physics}\ }\textbf {\bibinfo {volume} {63}},\
  \bibinfo {pages} {707} (\bibinfo {year} {1980})}\BibitemShut {NoStop}%
\bibitem [{\citenamefont {Al'tshuler}\ \emph {et~al.}(1981)\citenamefont
  {Al'tshuler}, \citenamefont {Aronov}, \citenamefont {Larkin},\ and\
  \citenamefont {Khmel'nitskii}}]{altshuler_anomalous_1981}%
  \BibitemOpen
  \bibfield  {author} {\bibinfo {author} {\bibfnamefont {B.~L.}\ \bibnamefont
  {Al'tshuler}}, \bibinfo {author} {\bibfnamefont {A.~G.}\ \bibnamefont
  {Aronov}}, \bibinfo {author} {\bibfnamefont {A.~I.}\ \bibnamefont {Larkin}},
  \ and\ \bibinfo {author} {\bibfnamefont {D.~E.}\ \bibnamefont
  {Khmel'nitskii}},\ }\href@noop {} {\bibfield  {journal} {\bibinfo  {journal}
  {Zh. Eksp. Teor. Fiz.}\ }\textbf {\bibinfo {volume} {81}},\ \bibinfo {pages}
  {768} (\bibinfo {year} {1981})}\BibitemShut {NoStop}%
\bibitem [{\citenamefont {Iordanskii}\ \emph {et~al.}(1994)\citenamefont
  {Iordanskii}, \citenamefont {Lyanda-Geller},\ and\ \citenamefont
  {Pikus}}]{iordanskii_weak_1994}%
  \BibitemOpen
  \bibfield  {author} {\bibinfo {author} {\bibfnamefont {S.~V.}\ \bibnamefont
  {Iordanskii}}, \bibinfo {author} {\bibfnamefont {Y.~B.}\ \bibnamefont
  {Lyanda-Geller}}, \ and\ \bibinfo {author} {\bibfnamefont {G.~E.}\
  \bibnamefont {Pikus}},\ }\href@noop {} {\bibfield  {journal} {\bibinfo
  {journal} {JETP Lett.}\ }\textbf {\bibinfo {volume} {60}},\ \bibinfo {pages}
  {206} (\bibinfo {year} {1994})}\BibitemShut {NoStop}%
\bibitem [{\citenamefont {Xu}\ \emph {et~al.}(2014)\citenamefont {Xu},
  \citenamefont {Phelan}, \citenamefont {Seok~Jeong}, \citenamefont
  {Andre~Mkhoyan},\ and\ \citenamefont {Jalan}}]{xu_stoichiometry-driven_2014}%
  \BibitemOpen
  \bibfield  {author} {\bibinfo {author} {\bibfnamefont {P.}~\bibnamefont
  {Xu}}, \bibinfo {author} {\bibfnamefont {D.}~\bibnamefont {Phelan}}, \bibinfo
  {author} {\bibfnamefont {J.}~\bibnamefont {Seok~Jeong}}, \bibinfo {author}
  {\bibfnamefont {K.}~\bibnamefont {Andre~Mkhoyan}}, \ and\ \bibinfo {author}
  {\bibfnamefont {B.}~\bibnamefont {Jalan}},\ }\href {\doibase
  10.1063/1.4866867} {\bibfield  {journal} {\bibinfo  {journal} {Applied
  Physics Letters}\ }\textbf {\bibinfo {volume} {104}},\ \bibinfo {pages}
  {082109} (\bibinfo {year} {2014})}\BibitemShut {NoStop}%
\bibitem [{\citenamefont {Xu}\ \emph {et~al.}(2016{\natexlab{a}})\citenamefont
  {Xu}, \citenamefont {Droubay}, \citenamefont {Jeong}, \citenamefont
  {Mkhoyan}, \citenamefont {Sushko}, \citenamefont {Chambers},\ and\
  \citenamefont {Jalan}}]{xu_quasi_2016}%
  \BibitemOpen
  \bibfield  {author} {\bibinfo {author} {\bibfnamefont {P.}~\bibnamefont
  {Xu}}, \bibinfo {author} {\bibfnamefont {T.~C.}\ \bibnamefont {Droubay}},
  \bibinfo {author} {\bibfnamefont {J.~S.}\ \bibnamefont {Jeong}}, \bibinfo
  {author} {\bibfnamefont {K.~A.}\ \bibnamefont {Mkhoyan}}, \bibinfo {author}
  {\bibfnamefont {P.~V.}\ \bibnamefont {Sushko}}, \bibinfo {author}
  {\bibfnamefont {S.~A.}\ \bibnamefont {Chambers}}, \ and\ \bibinfo {author}
  {\bibfnamefont {B.}~\bibnamefont {Jalan}},\ }\href {\doibase
  10.1002/admi.201500432} {\bibfield  {journal} {\bibinfo  {journal} {Advanced
  Materials Interfaces}\ }\textbf {\bibinfo {volume} {3}},\ \bibinfo {pages}
  {1500432} (\bibinfo {year} {2016}{\natexlab{a}})}\BibitemShut {NoStop}%
\bibitem [{\citenamefont {Xu}\ \emph {et~al.}(2016{\natexlab{b}})\citenamefont
  {Xu}, \citenamefont {Ayino}, \citenamefont {Cheng}, \citenamefont {Pribiag},
  \citenamefont {Comes}, \citenamefont {Sushko}, \citenamefont {Chambers},\
  and\ \citenamefont {Jalan}}]{xu_predictive_2016}%
  \BibitemOpen
  \bibfield  {author} {\bibinfo {author} {\bibfnamefont {P.}~\bibnamefont
  {Xu}}, \bibinfo {author} {\bibfnamefont {Y.}~\bibnamefont {Ayino}}, \bibinfo
  {author} {\bibfnamefont {C.}~\bibnamefont {Cheng}}, \bibinfo {author}
  {\bibfnamefont {V.~S.}\ \bibnamefont {Pribiag}}, \bibinfo {author}
  {\bibfnamefont {R.~B.}\ \bibnamefont {Comes}}, \bibinfo {author}
  {\bibfnamefont {P.~V.}\ \bibnamefont {Sushko}}, \bibinfo {author}
  {\bibfnamefont {S.~A.}\ \bibnamefont {Chambers}}, \ and\ \bibinfo {author}
  {\bibfnamefont {B.}~\bibnamefont {Jalan}},\ }\href {\doibase
  10.1103/PhysRevLett.117.106803} {\bibfield  {journal} {\bibinfo  {journal}
  {Physical Review Letters}\ }\textbf {\bibinfo {volume} {117}},\ \bibinfo
  {pages} {106803} (\bibinfo {year} {2016}{\natexlab{b}})}\BibitemShut
  {NoStop}%
\bibitem [{\citenamefont {Ohtomo}\ and\ \citenamefont
  {Hwang}(2004)}]{ohtomo_high-mobility_2004}%
  \BibitemOpen
  \bibfield  {author} {\bibinfo {author} {\bibfnamefont {A.}~\bibnamefont
  {Ohtomo}}\ and\ \bibinfo {author} {\bibfnamefont {H.~Y.}\ \bibnamefont
  {Hwang}},\ }\href {\doibase 10.1038/nature02308} {\bibfield  {journal}
  {\bibinfo  {journal} {Nature}\ }\textbf {\bibinfo {volume} {427}},\ \bibinfo
  {pages} {423} (\bibinfo {year} {2004})}\BibitemShut {NoStop}%
\bibitem [{\citenamefont {Takizawa}\ \emph {et~al.}(2006)\citenamefont
  {Takizawa}, \citenamefont {Wadati}, \citenamefont {Tanaka}, \citenamefont
  {Hashimoto}, \citenamefont {Yoshida}, \citenamefont {Fujimori}, \citenamefont
  {Chikamatsu}, \citenamefont {Kumigashira}, \citenamefont {Oshima},
  \citenamefont {Shibuya}, \citenamefont {Mihara}, \citenamefont {Ohnishi},
  \citenamefont {Lippmaa}, \citenamefont {Kawasaki}, \citenamefont {Koinuma},
  \citenamefont {Okamoto},\ and\ \citenamefont
  {Millis}}]{takizawa_photoemission_2006}%
  \BibitemOpen
  \bibfield  {author} {\bibinfo {author} {\bibfnamefont {M.}~\bibnamefont
  {Takizawa}}, \bibinfo {author} {\bibfnamefont {H.}~\bibnamefont {Wadati}},
  \bibinfo {author} {\bibfnamefont {K.}~\bibnamefont {Tanaka}}, \bibinfo
  {author} {\bibfnamefont {M.}~\bibnamefont {Hashimoto}}, \bibinfo {author}
  {\bibfnamefont {T.}~\bibnamefont {Yoshida}}, \bibinfo {author} {\bibfnamefont
  {A.}~\bibnamefont {Fujimori}}, \bibinfo {author} {\bibfnamefont
  {A.}~\bibnamefont {Chikamatsu}}, \bibinfo {author} {\bibfnamefont
  {H.}~\bibnamefont {Kumigashira}}, \bibinfo {author} {\bibfnamefont
  {M.}~\bibnamefont {Oshima}}, \bibinfo {author} {\bibfnamefont
  {K.}~\bibnamefont {Shibuya}}, \bibinfo {author} {\bibfnamefont
  {T.}~\bibnamefont {Mihara}}, \bibinfo {author} {\bibfnamefont
  {T.}~\bibnamefont {Ohnishi}}, \bibinfo {author} {\bibfnamefont
  {M.}~\bibnamefont {Lippmaa}}, \bibinfo {author} {\bibfnamefont
  {M.}~\bibnamefont {Kawasaki}}, \bibinfo {author} {\bibfnamefont
  {H.}~\bibnamefont {Koinuma}}, \bibinfo {author} {\bibfnamefont
  {S.}~\bibnamefont {Okamoto}}, \ and\ \bibinfo {author} {\bibfnamefont
  {A.~J.}\ \bibnamefont {Millis}},\ }\href {\doibase
  10.1103/PhysRevLett.97.057601} {\bibfield  {journal} {\bibinfo  {journal}
  {Physical Review Letters}\ }\textbf {\bibinfo {volume} {97}},\ \bibinfo
  {pages} {057601} (\bibinfo {year} {2006})}\BibitemShut {NoStop}%
\bibitem [{\citenamefont {Moetakef}\ \emph {et~al.}(2011)\citenamefont
  {Moetakef}, \citenamefont {Cain}, \citenamefont {Ouellette}, \citenamefont
  {Zhang}, \citenamefont {Klenov}, \citenamefont {Janotti}, \citenamefont
  {Van~de Walle}, \citenamefont {Rajan}, \citenamefont {Allen},\ and\
  \citenamefont {Stemmer}}]{moetakef_electrostatic_2011}%
  \BibitemOpen
  \bibfield  {author} {\bibinfo {author} {\bibfnamefont {P.}~\bibnamefont
  {Moetakef}}, \bibinfo {author} {\bibfnamefont {T.~A.}\ \bibnamefont {Cain}},
  \bibinfo {author} {\bibfnamefont {D.~G.}\ \bibnamefont {Ouellette}}, \bibinfo
  {author} {\bibfnamefont {J.~Y.}\ \bibnamefont {Zhang}}, \bibinfo {author}
  {\bibfnamefont {D.~O.}\ \bibnamefont {Klenov}}, \bibinfo {author}
  {\bibfnamefont {A.}~\bibnamefont {Janotti}}, \bibinfo {author} {\bibfnamefont
  {C.~G.}\ \bibnamefont {Van~de Walle}}, \bibinfo {author} {\bibfnamefont
  {S.}~\bibnamefont {Rajan}}, \bibinfo {author} {\bibfnamefont {S.~J.}\
  \bibnamefont {Allen}}, \ and\ \bibinfo {author} {\bibfnamefont
  {S.}~\bibnamefont {Stemmer}},\ }\href {\doibase 10.1063/1.3669402} {\bibfield
   {journal} {\bibinfo  {journal} {Applied Physics Letters}\ }\textbf {\bibinfo
  {volume} {99}},\ \bibinfo {pages} {232116} (\bibinfo {year}
  {2011})}\BibitemShut {NoStop}%
\bibitem [{\citenamefont {Sulpizio}\ \emph {et~al.}(2014)\citenamefont
  {Sulpizio}, \citenamefont {Ilani}, \citenamefont {Irvin},\ and\ \citenamefont
  {Levy}}]{sulpizio_nanoscale_2014}%
  \BibitemOpen
  \bibfield  {author} {\bibinfo {author} {\bibfnamefont {J.~A.}\ \bibnamefont
  {Sulpizio}}, \bibinfo {author} {\bibfnamefont {S.}~\bibnamefont {Ilani}},
  \bibinfo {author} {\bibfnamefont {P.}~\bibnamefont {Irvin}}, \ and\ \bibinfo
  {author} {\bibfnamefont {J.}~\bibnamefont {Levy}},\ }\href {\doibase
  10.1146/annurev-matsci-070813-113437} {\bibfield  {journal} {\bibinfo
  {journal} {Annual Review of Materials Research}\ }\textbf {\bibinfo {volume}
  {44}},\ \bibinfo {pages} {117} (\bibinfo {year} {2014})}\BibitemShut
  {NoStop}%
\bibitem [{\citenamefont {Reyren}\ \emph {et~al.}(2007)\citenamefont {Reyren},
  \citenamefont {Thiel}, \citenamefont {Caviglia}, \citenamefont {Kourkoutis},
  \citenamefont {Hammerl}, \citenamefont {Richter}, \citenamefont {Schneider},
  \citenamefont {Kopp}, \citenamefont {Ruetschi}, \citenamefont {Jaccard},
  \citenamefont {Gabay}, \citenamefont {Muller}, \citenamefont {Triscone},\
  and\ \citenamefont {Mannhart}}]{reyren_superconducting_2007}%
  \BibitemOpen
  \bibfield  {author} {\bibinfo {author} {\bibfnamefont {N.}~\bibnamefont
  {Reyren}}, \bibinfo {author} {\bibfnamefont {S.}~\bibnamefont {Thiel}},
  \bibinfo {author} {\bibfnamefont {A.~D.}\ \bibnamefont {Caviglia}}, \bibinfo
  {author} {\bibfnamefont {L.~F.}\ \bibnamefont {Kourkoutis}}, \bibinfo
  {author} {\bibfnamefont {G.}~\bibnamefont {Hammerl}}, \bibinfo {author}
  {\bibfnamefont {C.}~\bibnamefont {Richter}}, \bibinfo {author} {\bibfnamefont
  {C.~W.}\ \bibnamefont {Schneider}}, \bibinfo {author} {\bibfnamefont
  {T.}~\bibnamefont {Kopp}}, \bibinfo {author} {\bibfnamefont {A.-S.}\
  \bibnamefont {Ruetschi}}, \bibinfo {author} {\bibfnamefont {D.}~\bibnamefont
  {Jaccard}}, \bibinfo {author} {\bibfnamefont {M.}~\bibnamefont {Gabay}},
  \bibinfo {author} {\bibfnamefont {D.~A.}\ \bibnamefont {Muller}}, \bibinfo
  {author} {\bibfnamefont {J.-M.}\ \bibnamefont {Triscone}}, \ and\ \bibinfo
  {author} {\bibfnamefont {J.}~\bibnamefont {Mannhart}},\ }\href {\doibase
  10.1126/science.1146006} {\bibfield  {journal} {\bibinfo  {journal}
  {Science}\ }\textbf {\bibinfo {volume} {317}},\ \bibinfo {pages} {1196}
  (\bibinfo {year} {2007})}\BibitemShut {NoStop}%
\bibitem [{\citenamefont {Caviglia}\ \emph {et~al.}(2008)\citenamefont
  {Caviglia}, \citenamefont {Gariglio}, \citenamefont {Reyren}, \citenamefont
  {Jaccard}, \citenamefont {Schneider}, \citenamefont {Gabay}, \citenamefont
  {Thiel}, \citenamefont {Hammerl}, \citenamefont {Mannhart},\ and\
  \citenamefont {Triscone}}]{caviglia_electric_2008}%
  \BibitemOpen
  \bibfield  {author} {\bibinfo {author} {\bibfnamefont {A.~D.}\ \bibnamefont
  {Caviglia}}, \bibinfo {author} {\bibfnamefont {S.}~\bibnamefont {Gariglio}},
  \bibinfo {author} {\bibfnamefont {N.}~\bibnamefont {Reyren}}, \bibinfo
  {author} {\bibfnamefont {D.}~\bibnamefont {Jaccard}}, \bibinfo {author}
  {\bibfnamefont {T.}~\bibnamefont {Schneider}}, \bibinfo {author}
  {\bibfnamefont {M.}~\bibnamefont {Gabay}}, \bibinfo {author} {\bibfnamefont
  {S.}~\bibnamefont {Thiel}}, \bibinfo {author} {\bibfnamefont
  {G.}~\bibnamefont {Hammerl}}, \bibinfo {author} {\bibfnamefont
  {J.}~\bibnamefont {Mannhart}}, \ and\ \bibinfo {author} {\bibfnamefont
  {J.-M.}\ \bibnamefont {Triscone}},\ }\href {\doibase 10.1038/nature07576}
  {\bibfield  {journal} {\bibinfo  {journal} {Nature}\ }\textbf {\bibinfo
  {volume} {456}},\ \bibinfo {pages} {624} (\bibinfo {year}
  {2008})}\BibitemShut {NoStop}%
\bibitem [{\citenamefont {Biscaras}\ \emph {et~al.}(2010)\citenamefont
  {Biscaras}, \citenamefont {Bergeal}, \citenamefont {Kushwaha}, \citenamefont
  {Wolf}, \citenamefont {Rastogi}, \citenamefont {Budhani},\ and\ \citenamefont
  {Lesueur}}]{biscaras_two-dimensional_2010}%
  \BibitemOpen
  \bibfield  {author} {\bibinfo {author} {\bibfnamefont {J.}~\bibnamefont
  {Biscaras}}, \bibinfo {author} {\bibfnamefont {N.}~\bibnamefont {Bergeal}},
  \bibinfo {author} {\bibfnamefont {A.}~\bibnamefont {Kushwaha}}, \bibinfo
  {author} {\bibfnamefont {T.}~\bibnamefont {Wolf}}, \bibinfo {author}
  {\bibfnamefont {A.}~\bibnamefont {Rastogi}}, \bibinfo {author} {\bibfnamefont
  {R.}~\bibnamefont {Budhani}}, \ and\ \bibinfo {author} {\bibfnamefont
  {J.}~\bibnamefont {Lesueur}},\ }\href {\doibase 10.1038/ncomms1084}
  {\bibfield  {journal} {\bibinfo  {journal} {Nature Communications}\ }\textbf
  {\bibinfo {volume} {1}},\ \bibinfo {pages} {89} (\bibinfo {year}
  {2010})}\BibitemShut {NoStop}%
\bibitem [{\citenamefont {Ben~Shalom}\ \emph {et~al.}(2010)\citenamefont
  {Ben~Shalom}, \citenamefont {Sachs}, \citenamefont {Rakhmilevitch},
  \citenamefont {Palevski},\ and\ \citenamefont
  {Dagan}}]{ben_shalom_tuning_2010}%
  \BibitemOpen
  \bibfield  {author} {\bibinfo {author} {\bibfnamefont {M.}~\bibnamefont
  {Ben~Shalom}}, \bibinfo {author} {\bibfnamefont {M.}~\bibnamefont {Sachs}},
  \bibinfo {author} {\bibfnamefont {D.}~\bibnamefont {Rakhmilevitch}}, \bibinfo
  {author} {\bibfnamefont {A.}~\bibnamefont {Palevski}}, \ and\ \bibinfo
  {author} {\bibfnamefont {Y.}~\bibnamefont {Dagan}},\ }\href {\doibase
  10.1103/PhysRevLett.104.126802} {\bibfield  {journal} {\bibinfo  {journal}
  {Physical Review Letters}\ }\textbf {\bibinfo {volume} {104}},\ \bibinfo
  {pages} {126802} (\bibinfo {year} {2010})}\BibitemShut {NoStop}%
\bibitem [{\citenamefont {Caviglia}\ \emph {et~al.}(2010)\citenamefont
  {Caviglia}, \citenamefont {Gabay}, \citenamefont {Gariglio}, \citenamefont
  {Reyren}, \citenamefont {Cancellieri},\ and\ \citenamefont
  {Triscone}}]{caviglia_tunable_2010}%
  \BibitemOpen
  \bibfield  {author} {\bibinfo {author} {\bibfnamefont {A.~D.}\ \bibnamefont
  {Caviglia}}, \bibinfo {author} {\bibfnamefont {M.}~\bibnamefont {Gabay}},
  \bibinfo {author} {\bibfnamefont {S.}~\bibnamefont {Gariglio}}, \bibinfo
  {author} {\bibfnamefont {N.}~\bibnamefont {Reyren}}, \bibinfo {author}
  {\bibfnamefont {C.}~\bibnamefont {Cancellieri}}, \ and\ \bibinfo {author}
  {\bibfnamefont {J.-M.}\ \bibnamefont {Triscone}},\ }\href {\doibase
  10.1103/PhysRevLett.104.126803} {\bibfield  {journal} {\bibinfo  {journal}
  {Physical Review Letters}\ }\textbf {\bibinfo {volume} {104}},\ \bibinfo
  {pages} {126803} (\bibinfo {year} {2010})}\BibitemShut {NoStop}%
\bibitem [{\citenamefont {Bal}\ \emph {et~al.}(2018)\citenamefont {Bal},
  \citenamefont {Huang}, \citenamefont {Han}, \citenamefont {{Ariando}},
  \citenamefont {Venkatesan},\ and\ \citenamefont
  {Chandrasekhar}}]{bal_strong_2018}%
  \BibitemOpen
  \bibfield  {author} {\bibinfo {author} {\bibfnamefont {V.~V.}\ \bibnamefont
  {Bal}}, \bibinfo {author} {\bibfnamefont {Z.}~\bibnamefont {Huang}}, \bibinfo
  {author} {\bibfnamefont {K.}~\bibnamefont {Han}}, \bibinfo {author}
  {\bibnamefont {{Ariando}}}, \bibinfo {author} {\bibfnamefont
  {T.}~\bibnamefont {Venkatesan}}, \ and\ \bibinfo {author} {\bibfnamefont
  {V.}~\bibnamefont {Chandrasekhar}},\ }\href {\doibase
  10.1103/PhysRevB.98.085416} {\bibfield  {journal} {\bibinfo  {journal}
  {Physical Review B}\ }\textbf {\bibinfo {volume} {98}},\ \bibinfo {pages}
  {085416} (\bibinfo {year} {2018})}\BibitemShut {NoStop}%
\bibitem [{\citenamefont {Brinkman}\ \emph {et~al.}(2007)\citenamefont
  {Brinkman}, \citenamefont {Huijben}, \citenamefont {van Zalk}, \citenamefont
  {Huijben}, \citenamefont {Zeitler}, \citenamefont {Maan}, \citenamefont
  {van~der Wiel}, \citenamefont {Rijnders}, \citenamefont {Blank},\ and\
  \citenamefont {Hilgenkamp}}]{brinkman_magnetic_2007}%
  \BibitemOpen
  \bibfield  {author} {\bibinfo {author} {\bibfnamefont {A.}~\bibnamefont
  {Brinkman}}, \bibinfo {author} {\bibfnamefont {M.}~\bibnamefont {Huijben}},
  \bibinfo {author} {\bibfnamefont {M.}~\bibnamefont {van Zalk}}, \bibinfo
  {author} {\bibfnamefont {J.}~\bibnamefont {Huijben}}, \bibinfo {author}
  {\bibfnamefont {U.}~\bibnamefont {Zeitler}}, \bibinfo {author} {\bibfnamefont
  {J.~C.}\ \bibnamefont {Maan}}, \bibinfo {author} {\bibfnamefont {W.~G.}\
  \bibnamefont {van~der Wiel}}, \bibinfo {author} {\bibfnamefont
  {G.}~\bibnamefont {Rijnders}}, \bibinfo {author} {\bibfnamefont {D.~H.~A.}\
  \bibnamefont {Blank}}, \ and\ \bibinfo {author} {\bibfnamefont
  {H.}~\bibnamefont {Hilgenkamp}},\ }\href {\doibase 10.1038/nmat1931}
  {\bibfield  {journal} {\bibinfo  {journal} {Nature Materials}\ }\textbf
  {\bibinfo {volume} {6}},\ \bibinfo {pages} {493} (\bibinfo {year}
  {2007})}\BibitemShut {NoStop}%
\bibitem [{\citenamefont {Li}\ \emph {et~al.}(2011)\citenamefont {Li},
  \citenamefont {Richter}, \citenamefont {Mannhart},\ and\ \citenamefont
  {Ashoori}}]{li_coexistence_2011}%
  \BibitemOpen
  \bibfield  {author} {\bibinfo {author} {\bibfnamefont {L.}~\bibnamefont
  {Li}}, \bibinfo {author} {\bibfnamefont {C.}~\bibnamefont {Richter}},
  \bibinfo {author} {\bibfnamefont {J.}~\bibnamefont {Mannhart}}, \ and\
  \bibinfo {author} {\bibfnamefont {R.~C.}\ \bibnamefont {Ashoori}},\ }\href
  {\doibase 10.1038/nphys2080} {\bibfield  {journal} {\bibinfo  {journal}
  {Nature Physics}\ }\textbf {\bibinfo {volume} {7}},\ \bibinfo {pages} {762}
  (\bibinfo {year} {2011})}\BibitemShut {NoStop}%
\bibitem [{\citenamefont {Bert}\ \emph {et~al.}(2011)\citenamefont {Bert},
  \citenamefont {Kalisky}, \citenamefont {Bell}, \citenamefont {Kim},
  \citenamefont {Hikita}, \citenamefont {Hwang},\ and\ \citenamefont
  {Moler}}]{bert_direct_2011}%
  \BibitemOpen
  \bibfield  {author} {\bibinfo {author} {\bibfnamefont {J.~A.}\ \bibnamefont
  {Bert}}, \bibinfo {author} {\bibfnamefont {B.}~\bibnamefont {Kalisky}},
  \bibinfo {author} {\bibfnamefont {C.}~\bibnamefont {Bell}}, \bibinfo {author}
  {\bibfnamefont {M.}~\bibnamefont {Kim}}, \bibinfo {author} {\bibfnamefont
  {Y.}~\bibnamefont {Hikita}}, \bibinfo {author} {\bibfnamefont {H.~Y.}\
  \bibnamefont {Hwang}}, \ and\ \bibinfo {author} {\bibfnamefont {K.~A.}\
  \bibnamefont {Moler}},\ }\href {\doibase 10.1038/nphys2079} {\bibfield
  {journal} {\bibinfo  {journal} {Nature Physics}\ }\textbf {\bibinfo {volume}
  {7}},\ \bibinfo {pages} {767} (\bibinfo {year} {2011})}\BibitemShut {NoStop}%
\bibitem [{\citenamefont {Dikin}\ \emph {et~al.}(2011)\citenamefont {Dikin},
  \citenamefont {Mehta}, \citenamefont {Bark}, \citenamefont {Folkman},
  \citenamefont {Eom},\ and\ \citenamefont
  {Chandrasekhar}}]{dikin_coexistence_2011}%
  \BibitemOpen
  \bibfield  {author} {\bibinfo {author} {\bibfnamefont {D.~A.}\ \bibnamefont
  {Dikin}}, \bibinfo {author} {\bibfnamefont {M.}~\bibnamefont {Mehta}},
  \bibinfo {author} {\bibfnamefont {C.~W.}\ \bibnamefont {Bark}}, \bibinfo
  {author} {\bibfnamefont {C.~M.}\ \bibnamefont {Folkman}}, \bibinfo {author}
  {\bibfnamefont {C.~B.}\ \bibnamefont {Eom}}, \ and\ \bibinfo {author}
  {\bibfnamefont {V.}~\bibnamefont {Chandrasekhar}},\ }\href {\doibase
  10.1103/PhysRevLett.107.056802} {\bibfield  {journal} {\bibinfo  {journal}
  {Physical Review Letters}\ }\textbf {\bibinfo {volume} {107}},\ \bibinfo
  {pages} {056802} (\bibinfo {year} {2011})}\BibitemShut {NoStop}%
\bibitem [{\citenamefont {Moetakef}\ \emph {et~al.}(2012)\citenamefont
  {Moetakef}, \citenamefont {Williams}, \citenamefont {Ouellette},
  \citenamefont {Kajdos}, \citenamefont {Goldhaber-Gordon}, \citenamefont
  {Allen},\ and\ \citenamefont {Stemmer}}]{moetakef_carrier-controlled_2012}%
  \BibitemOpen
  \bibfield  {author} {\bibinfo {author} {\bibfnamefont {P.}~\bibnamefont
  {Moetakef}}, \bibinfo {author} {\bibfnamefont {J.~R.}\ \bibnamefont
  {Williams}}, \bibinfo {author} {\bibfnamefont {D.~G.}\ \bibnamefont
  {Ouellette}}, \bibinfo {author} {\bibfnamefont {A.~P.}\ \bibnamefont
  {Kajdos}}, \bibinfo {author} {\bibfnamefont {D.}~\bibnamefont
  {Goldhaber-Gordon}}, \bibinfo {author} {\bibfnamefont {S.~J.}\ \bibnamefont
  {Allen}}, \ and\ \bibinfo {author} {\bibfnamefont {S.}~\bibnamefont
  {Stemmer}},\ }\href {\doibase 10.1103/PhysRevX.2.021014} {\bibfield
  {journal} {\bibinfo  {journal} {Physical Review X}\ }\textbf {\bibinfo
  {volume} {2}},\ \bibinfo {pages} {021014} (\bibinfo {year}
  {2012})}\BibitemShut {NoStop}%
\bibitem [{\citenamefont {Joshua}\ \emph {et~al.}(2013)\citenamefont {Joshua},
  \citenamefont {Ruhman}, \citenamefont {Pecker}, \citenamefont {Altman},\ and\
  \citenamefont {Ilani}}]{joshua_gate-tunable_2013}%
  \BibitemOpen
  \bibfield  {author} {\bibinfo {author} {\bibfnamefont {A.}~\bibnamefont
  {Joshua}}, \bibinfo {author} {\bibfnamefont {J.}~\bibnamefont {Ruhman}},
  \bibinfo {author} {\bibfnamefont {S.}~\bibnamefont {Pecker}}, \bibinfo
  {author} {\bibfnamefont {E.}~\bibnamefont {Altman}}, \ and\ \bibinfo {author}
  {\bibfnamefont {S.}~\bibnamefont {Ilani}},\ }\href {\doibase
  10.1073/pnas.1221453110} {\bibfield  {journal} {\bibinfo  {journal}
  {Proceedings of the National Academy of Sciences}\ }\textbf {\bibinfo
  {volume} {110}},\ \bibinfo {pages} {9633} (\bibinfo {year}
  {2013})}\BibitemShut {NoStop}%
\bibitem [{\citenamefont {Anahory}\ \emph {et~al.}(2016)\citenamefont
  {Anahory}, \citenamefont {Embon}, \citenamefont {Li}, \citenamefont
  {Banerjee}, \citenamefont {Meltzer}, \citenamefont {Naren}, \citenamefont
  {Yakovenko}, \citenamefont {Cuppens}, \citenamefont {Myasoedov},
  \citenamefont {Rappaport}, \citenamefont {Huber}, \citenamefont {Michaeli},
  \citenamefont {Venkatesan}, \citenamefont {{Ariando}},\ and\ \citenamefont
  {Zeldov}}]{anahory_emergent_2016}%
  \BibitemOpen
  \bibfield  {author} {\bibinfo {author} {\bibfnamefont {Y.}~\bibnamefont
  {Anahory}}, \bibinfo {author} {\bibfnamefont {L.}~\bibnamefont {Embon}},
  \bibinfo {author} {\bibfnamefont {C.~J.}\ \bibnamefont {Li}}, \bibinfo
  {author} {\bibfnamefont {S.}~\bibnamefont {Banerjee}}, \bibinfo {author}
  {\bibfnamefont {A.}~\bibnamefont {Meltzer}}, \bibinfo {author} {\bibfnamefont
  {H.~R.}\ \bibnamefont {Naren}}, \bibinfo {author} {\bibfnamefont
  {A.}~\bibnamefont {Yakovenko}}, \bibinfo {author} {\bibfnamefont
  {J.}~\bibnamefont {Cuppens}}, \bibinfo {author} {\bibfnamefont
  {Y.}~\bibnamefont {Myasoedov}}, \bibinfo {author} {\bibfnamefont {M.~L.}\
  \bibnamefont {Rappaport}}, \bibinfo {author} {\bibfnamefont {M.~E.}\
  \bibnamefont {Huber}}, \bibinfo {author} {\bibfnamefont {K.}~\bibnamefont
  {Michaeli}}, \bibinfo {author} {\bibfnamefont {T.}~\bibnamefont
  {Venkatesan}}, \bibinfo {author} {\bibnamefont {{Ariando}}}, \ and\ \bibinfo
  {author} {\bibfnamefont {E.}~\bibnamefont {Zeldov}},\ }\href {\doibase
  10.1038/ncomms12566} {\bibfield  {journal} {\bibinfo  {journal} {Nature
  Communications}\ }\textbf {\bibinfo {volume} {7}},\ \bibinfo {pages} {12566}
  (\bibinfo {year} {2016})}\BibitemShut {NoStop}%
\bibitem [{\citenamefont {Ayino}\ \emph {et~al.}(2018)\citenamefont {Ayino},
  \citenamefont {Xu}, \citenamefont {Tigre-Lazo}, \citenamefont {Yue},
  \citenamefont {Jalan},\ and\ \citenamefont
  {Pribiag}}]{ayino_ferromagnetism_2018}%
  \BibitemOpen
  \bibfield  {author} {\bibinfo {author} {\bibfnamefont {Y.}~\bibnamefont
  {Ayino}}, \bibinfo {author} {\bibfnamefont {P.}~\bibnamefont {Xu}}, \bibinfo
  {author} {\bibfnamefont {J.}~\bibnamefont {Tigre-Lazo}}, \bibinfo {author}
  {\bibfnamefont {J.}~\bibnamefont {Yue}}, \bibinfo {author} {\bibfnamefont
  {B.}~\bibnamefont {Jalan}}, \ and\ \bibinfo {author} {\bibfnamefont {V.~S.}\
  \bibnamefont {Pribiag}},\ }\href {\doibase 10.1103/PhysRevMaterials.2.031401}
  {\bibfield  {journal} {\bibinfo  {journal} {Physical Review Materials}\
  }\textbf {\bibinfo {volume} {2}},\ \bibinfo {pages} {031401} (\bibinfo {year}
  {2018})}\BibitemShut {NoStop}%
\bibitem [{\citenamefont {Amow}\ and\ \citenamefont
  {Greedan}(1996)}]{amow_structural_1996}%
  \BibitemOpen
  \bibfield  {author} {\bibinfo {author} {\bibfnamefont {G.}~\bibnamefont
  {Amow}}\ and\ \bibinfo {author} {\bibfnamefont {J.}~\bibnamefont {Greedan}},\
  }\href {\doibase 10.1006/jssc.1996.0061} {\bibfield  {journal} {\bibinfo
  {journal} {Journal of Solid State Chemistry}\ }\textbf {\bibinfo {volume}
  {121}},\ \bibinfo {pages} {443} (\bibinfo {year} {1996})}\BibitemShut
  {NoStop}%
\bibitem [{\citenamefont {Sefat}\ \emph {et~al.}(2006)\citenamefont {Sefat},
  \citenamefont {Greedan},\ and\ \citenamefont
  {Cranswick}}]{sefat_effect_2006}%
  \BibitemOpen
  \bibfield  {author} {\bibinfo {author} {\bibfnamefont {A.~S.}\ \bibnamefont
  {Sefat}}, \bibinfo {author} {\bibfnamefont {J.~E.}\ \bibnamefont {Greedan}},
  \ and\ \bibinfo {author} {\bibfnamefont {L.}~\bibnamefont {Cranswick}},\
  }\href {\doibase 10.1103/PhysRevB.74.104418} {\bibfield  {journal} {\bibinfo
  {journal} {Physical Review B}\ }\textbf {\bibinfo {volume} {74}},\ \bibinfo
  {pages} {104418} (\bibinfo {year} {2006})}\BibitemShut {NoStop}%
\bibitem [{\citenamefont {Jalan}\ \emph {et~al.}(2009)\citenamefont {Jalan},
  \citenamefont {Moetakef},\ and\ \citenamefont
  {Stemmer}}]{jalan_molecular_2009}%
  \BibitemOpen
  \bibfield  {author} {\bibinfo {author} {\bibfnamefont {B.}~\bibnamefont
  {Jalan}}, \bibinfo {author} {\bibfnamefont {P.}~\bibnamefont {Moetakef}}, \
  and\ \bibinfo {author} {\bibfnamefont {S.}~\bibnamefont {Stemmer}},\ }\href
  {\doibase 10.1063/1.3184767} {\bibfield  {journal} {\bibinfo  {journal}
  {Applied Physics Letters}\ }\textbf {\bibinfo {volume} {95}},\ \bibinfo
  {pages} {032906} (\bibinfo {year} {2009})}\BibitemShut {NoStop}%
\bibitem [{\citenamefont {Janotti}\ \emph {et~al.}(2012)\citenamefont
  {Janotti}, \citenamefont {Bjaalie}, \citenamefont {Gordon},\ and\
  \citenamefont {Van~de Walle}}]{janotti_controlling_2012}%
  \BibitemOpen
  \bibfield  {author} {\bibinfo {author} {\bibfnamefont {A.}~\bibnamefont
  {Janotti}}, \bibinfo {author} {\bibfnamefont {L.}~\bibnamefont {Bjaalie}},
  \bibinfo {author} {\bibfnamefont {L.}~\bibnamefont {Gordon}}, \ and\ \bibinfo
  {author} {\bibfnamefont {C.~G.}\ \bibnamefont {Van~de Walle}},\ }\href
  {\doibase 10.1103/PhysRevB.86.241108} {\bibfield  {journal} {\bibinfo
  {journal} {Physical Review B}\ }\textbf {\bibinfo {volume} {86}} (\bibinfo
  {year} {2012}),\ 10.1103/PhysRevB.86.241108}\BibitemShut {NoStop}%
\bibitem [{\citenamefont {Minkov}\ \emph {et~al.}(2004)\citenamefont {Minkov},
  \citenamefont {Germanenko}, \citenamefont {Rut}, \citenamefont
  {Sherstobitov}, \citenamefont {Golub}, \citenamefont {Zvonkov},\ and\
  \citenamefont {Willander}}]{minkov_weak_2004}%
  \BibitemOpen
  \bibfield  {author} {\bibinfo {author} {\bibfnamefont {G.~M.}\ \bibnamefont
  {Minkov}}, \bibinfo {author} {\bibfnamefont {A.~V.}\ \bibnamefont
  {Germanenko}}, \bibinfo {author} {\bibfnamefont {O.~E.}\ \bibnamefont {Rut}},
  \bibinfo {author} {\bibfnamefont {A.~A.}\ \bibnamefont {Sherstobitov}},
  \bibinfo {author} {\bibfnamefont {L.~E.}\ \bibnamefont {Golub}}, \bibinfo
  {author} {\bibfnamefont {B.~N.}\ \bibnamefont {Zvonkov}}, \ and\ \bibinfo
  {author} {\bibfnamefont {M.}~\bibnamefont {Willander}},\ }\href {\doibase
  10.1103/PhysRevB.70.155323} {\bibfield  {journal} {\bibinfo  {journal} {Phys.
  Rev. B}\ }\textbf {\bibinfo {volume} {70}},\ \bibinfo {pages} {155323}
  (\bibinfo {year} {2004})}\BibitemShut {NoStop}%
\bibitem [{\citenamefont {Knap}\ \emph {et~al.}(1996)\citenamefont {Knap},
  \citenamefont {Skierbiszewski}, \citenamefont {Zduniak}, \citenamefont
  {Litwin-Staszewska}, \citenamefont {Bertho}, \citenamefont {Kobbi},
  \citenamefont {Robert}, \citenamefont {Pikus}, \citenamefont {Pikus},\ and\
  \citenamefont {Iordanskii}}]{knap_weak_1996}%
  \BibitemOpen
  \bibfield  {author} {\bibinfo {author} {\bibfnamefont {W.}~\bibnamefont
  {Knap}}, \bibinfo {author} {\bibfnamefont {C.}~\bibnamefont
  {Skierbiszewski}}, \bibinfo {author} {\bibfnamefont {A.}~\bibnamefont
  {Zduniak}}, \bibinfo {author} {\bibfnamefont {E.}~\bibnamefont
  {Litwin-Staszewska}}, \bibinfo {author} {\bibfnamefont {D.}~\bibnamefont
  {Bertho}}, \bibinfo {author} {\bibfnamefont {F.}~\bibnamefont {Kobbi}},
  \bibinfo {author} {\bibfnamefont {J.~L.}\ \bibnamefont {Robert}}, \bibinfo
  {author} {\bibfnamefont {G.~E.}\ \bibnamefont {Pikus}}, \bibinfo {author}
  {\bibfnamefont {F.~G.}\ \bibnamefont {Pikus}}, \ and\ \bibinfo {author}
  {\bibfnamefont {S.~V.}\ \bibnamefont {Iordanskii}},\ }\href@noop {}
  {\bibfield  {journal} {\bibinfo  {journal} {Physical Review B}\ }\textbf
  {\bibinfo {volume} {53}},\ \bibinfo {pages} {3912} (\bibinfo {year}
  {1996})}\BibitemShut {NoStop}%
\bibitem [{Sup()}]{Supinfo}%
  \BibitemOpen
  \href@noop {} {\bibinfo  {journal} {Supplemental Material}\ }\BibitemShut
  {NoStop}%
\bibitem [{\citenamefont {Jeong}\ \emph {et~al.}(2016)\citenamefont {Jeong},
  \citenamefont {Topsakal}, \citenamefont {Xu}, \citenamefont {Jalan},
  \citenamefont {Wentzcovitch},\ and\ \citenamefont
  {Mkhoyan}}]{jeong_new_2016}%
  \BibitemOpen
\bibfield  {journal} {  }\bibfield  {author} {\bibinfo {author} {\bibfnamefont
  {J.~S.}\ \bibnamefont {Jeong}}, \bibinfo {author} {\bibfnamefont
  {M.}~\bibnamefont {Topsakal}}, \bibinfo {author} {\bibfnamefont
  {P.}~\bibnamefont {Xu}}, \bibinfo {author} {\bibfnamefont {B.}~\bibnamefont
  {Jalan}}, \bibinfo {author} {\bibfnamefont {R.~M.}\ \bibnamefont
  {Wentzcovitch}}, \ and\ \bibinfo {author} {\bibfnamefont {K.~A.}\
  \bibnamefont {Mkhoyan}},\ }\href {\doibase 10.1021/acs.nanolett.6b02532}
  {\bibfield  {journal} {\bibinfo  {journal} {Nano Letters}\ }\textbf {\bibinfo
  {volume} {16}},\ \bibinfo {pages} {6816} (\bibinfo {year}
  {2016})}\BibitemShut {NoStop}%
\bibitem [{\citenamefont {Nakamura}\ \emph {et~al.}(2012)\citenamefont
  {Nakamura}, \citenamefont {Koga},\ and\ \citenamefont
  {Kimura}}]{nakamura_experimental_2012}%
  \BibitemOpen
  \bibfield  {author} {\bibinfo {author} {\bibfnamefont {H.}~\bibnamefont
  {Nakamura}}, \bibinfo {author} {\bibfnamefont {T.}~\bibnamefont {Koga}}, \
  and\ \bibinfo {author} {\bibfnamefont {T.}~\bibnamefont {Kimura}},\ }\href
  {\doibase 10.1103/PhysRevLett.108.206601} {\bibfield  {journal} {\bibinfo
  {journal} {Physical Review Letters}\ }\textbf {\bibinfo {volume} {108}},\
  \bibinfo {pages} {206601} (\bibinfo {year} {2012})}\BibitemShut {NoStop}%
\bibitem [{\citenamefont {Mattheiss}(1972)}]{mattheiss_effect_1972-1}%
  \BibitemOpen
  \bibfield  {author} {\bibinfo {author} {\bibfnamefont {L.~F.}\ \bibnamefont
  {Mattheiss}},\ }\href {\doibase 10.1103/PhysRevB.6.4740} {\bibfield
  {journal} {\bibinfo  {journal} {Physical Review B}\ }\textbf {\bibinfo
  {volume} {6}},\ \bibinfo {pages} {4740} (\bibinfo {year} {1972})}\BibitemShut
  {NoStop}%
\bibitem [{\citenamefont {Maekawa}\ and\ \citenamefont
  {Fukuyama}(1981)}]{maekawa_magnetoresistance_1981}%
  \BibitemOpen
  \bibfield  {author} {\bibinfo {author} {\bibfnamefont {S.}~\bibnamefont
  {Maekawa}}\ and\ \bibinfo {author} {\bibfnamefont {H.}~\bibnamefont
  {Fukuyama}},\ }\href@noop {} {\bibfield  {journal} {\bibinfo  {journal} {J.
  Phys. Soc. Jpn.}\ }\textbf {\bibinfo {volume} {50}},\ \bibinfo {pages} {2516}
  (\bibinfo {year} {1981})}\BibitemShut {NoStop}%
\bibitem [{\citenamefont {Mal’shukov}\ \emph {et~al.}(1997)\citenamefont
  {Mal’shukov}, \citenamefont {Chao},\ and\ \citenamefont
  {Willander}}]{malshukov_magnetoresistance_1997}%
  \BibitemOpen
  \bibfield  {author} {\bibinfo {author} {\bibfnamefont {A.~G.}\ \bibnamefont
  {Mal’shukov}}, \bibinfo {author} {\bibfnamefont {K.~A.}\ \bibnamefont
  {Chao}}, \ and\ \bibinfo {author} {\bibfnamefont {M.}~\bibnamefont
  {Willander}},\ }\href {\doibase 10.1103/PhysRevB.56.6436} {\bibfield
  {journal} {\bibinfo  {journal} {Physical Review B}\ }\textbf {\bibinfo
  {volume} {56}},\ \bibinfo {pages} {6436} (\bibinfo {year}
  {1997})}\BibitemShut {NoStop}%
\bibitem [{\citenamefont {Zumbühl}\ \emph {et~al.}(2002)\citenamefont
  {Zumbühl}, \citenamefont {Miller}, \citenamefont {Marcus}, \citenamefont
  {Campman},\ and\ \citenamefont {Gossard}}]{zumbuhl_spin-orbit_2002}%
  \BibitemOpen
  \bibfield  {author} {\bibinfo {author} {\bibfnamefont {D.~M.}\ \bibnamefont
  {Zumbühl}}, \bibinfo {author} {\bibfnamefont {J.~B.}\ \bibnamefont
  {Miller}}, \bibinfo {author} {\bibfnamefont {C.~M.}\ \bibnamefont {Marcus}},
  \bibinfo {author} {\bibfnamefont {K.}~\bibnamefont {Campman}}, \ and\
  \bibinfo {author} {\bibfnamefont {A.~C.}\ \bibnamefont {Gossard}},\ }\href
  {\doibase 10.1103/PhysRevLett.89.276803} {\bibfield  {journal} {\bibinfo
  {journal} {Physical Review Letters}\ }\textbf {\bibinfo {volume} {89}},\
  \bibinfo {pages} {276803} (\bibinfo {year} {2002})}\BibitemShut {NoStop}%
\bibitem [{\citenamefont {Mathur}\ and\ \citenamefont
  {Baranger}(2001)}]{mathur_random_2001}%
  \BibitemOpen
  \bibfield  {author} {\bibinfo {author} {\bibfnamefont {H.}~\bibnamefont
  {Mathur}}\ and\ \bibinfo {author} {\bibfnamefont {H.~U.}\ \bibnamefont
  {Baranger}},\ }\href {\doibase 10.1103/PhysRevB.64.235325} {\bibfield
  {journal} {\bibinfo  {journal} {Physical Review B}\ }\textbf {\bibinfo
  {volume} {64}},\ \bibinfo {pages} {235325} (\bibinfo {year}
  {2001})}\BibitemShut {NoStop}%
\bibitem [{\citenamefont {Meyer}\ \emph {et~al.}(2002)\citenamefont {Meyer},
  \citenamefont {Altland},\ and\ \citenamefont
  {Altshuler}}]{meyer_quantum_2002}%
  \BibitemOpen
  \bibfield  {author} {\bibinfo {author} {\bibfnamefont {J.~S.}\ \bibnamefont
  {Meyer}}, \bibinfo {author} {\bibfnamefont {A.}~\bibnamefont {Altland}}, \
  and\ \bibinfo {author} {\bibfnamefont {B.~L.}\ \bibnamefont {Altshuler}},\
  }\href {\doibase 10.1103/PhysRevLett.89.206601} {\bibfield  {journal}
  {\bibinfo  {journal} {Physical Review Letters}\ }\textbf {\bibinfo {volume}
  {89}},\ \bibinfo {pages} {206601} (\bibinfo {year} {2002})}\BibitemShut
  {NoStop}%
\bibitem [{\citenamefont {Raichev}\ and\ \citenamefont
  {Vasilopoulos}(2000)}]{raichev_weak-localization_2000}%
  \BibitemOpen
  \bibfield  {author} {\bibinfo {author} {\bibfnamefont {O.~E.}\ \bibnamefont
  {Raichev}}\ and\ \bibinfo {author} {\bibfnamefont {P.}~\bibnamefont
  {Vasilopoulos}},\ }\href {\doibase 10.1088/0953-8984/12/5/307} {\bibfield
  {journal} {\bibinfo  {journal} {Journal of Physics: Condensed Matter}\
  }\textbf {\bibinfo {volume} {12}},\ \bibinfo {pages} {589} (\bibinfo {year}
  {2000})}\BibitemShut {NoStop}%
\bibitem [{\citenamefont {O'Handley}(2000)}]{ohandley_modern_2000}%
  \BibitemOpen
  \bibfield  {author} {\bibinfo {author} {\bibfnamefont {R.~C.}\ \bibnamefont
  {O'Handley}},\ }\href@noop {} {\emph {\bibinfo {title} {Modern {Magnetic}
  {Materials}: {Principles} and {Applications}}}}\ (\bibinfo  {publisher}
  {Wiley-Interscience},\ \bibinfo {year} {2000})\BibitemShut {NoStop}%
\bibitem [{\citenamefont {Dugaev}\ \emph {et~al.}(2001)\citenamefont {Dugaev},
  \citenamefont {Bruno},\ and\ \citenamefont {Barna\ifmmode~\acute{s}\else
  \'{s}\fi{}}}]{dugaev_weak_2001}%
  \BibitemOpen
  \bibfield  {author} {\bibinfo {author} {\bibfnamefont {V.~K.}\ \bibnamefont
  {Dugaev}}, \bibinfo {author} {\bibfnamefont {P.}~\bibnamefont {Bruno}}, \
  and\ \bibinfo {author} {\bibfnamefont {J.}~\bibnamefont
  {Barna\ifmmode~\acute{s}\else \'{s}\fi{}}},\ }\href {\doibase
  10.1103/PhysRevB.64.144423} {\bibfield  {journal} {\bibinfo  {journal} {Phys.
  Rev. B}\ }\textbf {\bibinfo {volume} {64}},\ \bibinfo {pages} {144423}
  (\bibinfo {year} {2001})}\BibitemShut {NoStop}%
\bibitem [{\citenamefont {Allia}\ \emph {et~al.}(2001)\citenamefont {Allia},
  \citenamefont {Coisson}, \citenamefont {Tiberto}, \citenamefont {Vinai},
  \citenamefont {Knobel}, \citenamefont {Novak},\ and\ \citenamefont
  {Nunes}}]{allia_granular_2001}%
  \BibitemOpen
  \bibfield  {author} {\bibinfo {author} {\bibfnamefont {P.}~\bibnamefont
  {Allia}}, \bibinfo {author} {\bibfnamefont {M.}~\bibnamefont {Coisson}},
  \bibinfo {author} {\bibfnamefont {P.}~\bibnamefont {Tiberto}}, \bibinfo
  {author} {\bibfnamefont {F.}~\bibnamefont {Vinai}}, \bibinfo {author}
  {\bibfnamefont {M.}~\bibnamefont {Knobel}}, \bibinfo {author} {\bibfnamefont
  {M.}~\bibnamefont {Novak}}, \ and\ \bibinfo {author} {\bibfnamefont
  {W.}~\bibnamefont {Nunes}},\ }\href {\doibase 10.1103/PhysRevB.64.144420}
  {\bibfield  {journal} {\bibinfo  {journal} {Physical Review B}\ }\textbf
  {\bibinfo {volume} {64}},\ \bibinfo {pages} {144420} (\bibinfo {year}
  {2001})}\BibitemShut {NoStop}%
\bibitem [{\citenamefont {Pentcheva}\ and\ \citenamefont
  {Pickett}(2006)}]{pentcheva_charge_2006}%
  \BibitemOpen
  \bibfield  {author} {\bibinfo {author} {\bibfnamefont {R.}~\bibnamefont
  {Pentcheva}}\ and\ \bibinfo {author} {\bibfnamefont {W.~E.}\ \bibnamefont
  {Pickett}},\ }\href {\doibase 10.1103/PhysRevB.74.035112} {\bibfield
  {journal} {\bibinfo  {journal} {Physical Review B}\ }\textbf {\bibinfo
  {volume} {74}},\ \bibinfo {pages} {035112} (\bibinfo {year}
  {2006})}\BibitemShut {NoStop}%
\end{thebibliography}%

\end{document}